\documentclass[12pt]{article}

\usepackage{latexsym}
\usepackage{amssymb,amsfonts,amsmath}
\usepackage{graphicx} 
\usepackage{indentfirst}
\usepackage{bbm}
\usepackage{amssymb}
\usepackage{verbatim}
\usepackage{amsmath, amsthm,amssymb}
\usepackage{mathrsfs}
\usepackage{hyperref}
\usepackage{amsfonts}
\usepackage{dsfont}
\usepackage{cite}
\usepackage{xcolor}
\usepackage[multiple]{footmisc}

\usepackage{braket}

\parskip=\medskipamount

\newcommand {\cC}{{\cal C}}

\newcommand {\cJ}{{\cal J}}

\newcommand {\cN}{{\cal N}}

\def\a{\alpha}

\def\k{\kappa}
\def\l{\lambda}

\def\s{\sigma}

\def\L{\Lambda}

\def\ri{{\rm i}}

\newcommand{\ad}{{\dot{\alpha}}}                           
\newcommand{\ve}{\varepsilon}                            

\newcommand{\hf}{\frac12}

\newcommand{\be}{\begin{equation}}
\newcommand{\ee}{\end{equation}}
\newcommand{\bea}{\begin{eqnarray}}
\newcommand{\eea}{\end{eqnarray}}

\def\double #1{#1{\hbox{\kern-2pt $#1$}}}

\newif\ifdtup

\newcommand{\bsubeq}{\begin{subequations}}
\newcommand{\esubeq}{\end{subequations}}

\numberwithin{equation}{section}

\newcommand{\sSU}{\mathsf{SU}}

\newcommand{\sISO}{\mathsf{ISO}}

\begin{document}

\begin{titlepage}
\begin{flushright}
October, 2020 \\
Revised version: November, 2020
\end{flushright}
\vspace{5mm}

\begin{center}
{\Large \bf 
Massless particles in five and higher dimensions}
\end{center}

\begin{center}

{\bf Sergei M. Kuzenko and Alec E. Pindur} \\
\vspace{5mm}

\footnotesize{
{\it Department of Physics M013, The University of Western Australia\\
35 Stirling Highway, Perth W.A. 6009, Australia}}
~\\
\vspace{2mm}
Email: \texttt{ 
sergei.kuzenko@uwa.edu.au, 21504287@student.uwa.edu.au}\\

\end{center}

\begin{abstract}
\baselineskip=14pt
We describe a five-dimensional analogue of Wigner's  operator equation 
${\mathbb W}_a = \lambda  P_a$, where ${\mathbb W}_a $ is the Pauli-Lubanski vector, 
$P_a$ the energy-momentum operator, and $\l$ the helicity of a massless particle. 
Higher dimensional generalisations are also given.
\end{abstract}
\vspace{5mm}

\vfill

\vfill
\end{titlepage}

\newpage
\renewcommand{\thefootnote}{\arabic{footnote}}
\setcounter{footnote}{0}

%

\allowdisplaybreaks

\section{Introduction}

The unitary representations of the Poincar\'e group in four dimensions were classified 
by Wigner in 1939 \cite{Wigner}, see \cite{Weinberg95} for a recent review. Our modern understanding of elementary particles is based on this classification. 

Unitary representations of the Poincar\'e group $\sISO_0(d-1, 1)$ in higher dimensions,
$d>4$, have been studied in the literature, see \cite{BB02,BB03,BB} and references therein.
However, significant interest in the topic remains due to some unexplored generalisations (including the supersymmetric case). It suffices to mention
the recent work by Weinberg \cite{Weinberg2020} in which
free massless field equations were derived
as an incidental consequence of the condition that the invariant Abelian
subgroup of the little group is represented trivially on states
of a single massless particle.
Unlike the analyses in \cite{BB02,BB03,BB,Weinberg2020},
in this paper we do not discuss the massless field equations in higher dimensions. 
We analyse covariant operator equations that correspond to  the 
irreducible massless representations of $\sISO_0(d-1,1)$ with a finite (discrete) spin.

We recall that the Poincar\'e algebra $\mathfrak{iso} (d-1,1) $  in $d$ dimensions is characterised  by the commutation relations\footnote{We make use of the mostly plus  Minkowski metric $\eta_{ab}$ and normalise the Levi-Civita tensor $\ve_{a_1 \dots a_d}$ 
by $\ve_{0 1\dots d-1}=1$.} 
\begin{subequations} \label{PA}
\bea
 \big[P_{a},  P_{b}\big] & = & 0 ~, \\
 \big[J_{ab},  P_{c}\big]  & = & \ri\eta_{ac}P_{b} - \ri\eta_{bc}P_a ~,\\
 \big[J_{a b}, J_{c d}\big]  & = &  \ri \eta_{a c}J_{b d} - \ri \eta_{a d} J_{b c} + \ri \eta_{b d} J_{a c} - \ri \eta_{bc} J_{a d} ~.
\eea
\end{subequations}
In any unitary representation of (the universal covering group of) the Poincar\'e group, 
the energy-momentum operator
$P_a$ and the Lorentz generators $J_{ab}$ are Hermitian.
For every dimension $d$, the operator $P^a P_a$ is a Casimir operator. Other Casimir operators are dimension dependent. 

In four dimensions, the second Casimir operator is 
${\mathbb W}^a{\mathbb W}_a $,
where 
\begin{equation}
{\mathbb W}^a=\frac12 \ve^{abcd} { J}_{bc}{ P}_d
\label{PLv}
\end{equation}
is the Pauli-Lubanski vector.
Using the commutation relations 
 \eqref{PA}, it follows that 
 the Pauli-Lubanski vector is translationally invariant, 
\begin{subequations}
\bea
\big[ P_a, {\mathbb W}_b \big]=0~,
\eea
and possesses the following properties:
\bea
{\mathbb W}^a{ P}_a&=&0 ~,  \\
\big[{ J}_{ab},{\mathbb W}_c \big]&=&\ri \eta_{ac}{\mathbb W}_b
-\ri  \eta_{bc}{\mathbb W}_a~,
\\ 
\big[ {\mathbb W}_a,{\mathbb W}_b \big]&=& \ri  \ve_{abcd}{\mathbb W}^c{P}^d~. 
\eea
\end{subequations}
The irreducible massive representations are characterised by the conditions 
\begin{subequations}
\bea
 { P}^a{ P}_a &=&-m^2\,{\mathbbm 1}~, \qquad m^2 > 0~, 
 \qquad 
 {\rm sign } \,{ P}^0 >0~, \label{1.4a}  \\
 {\mathbb W}^a{\mathbb W}_a &=& m^2s(s+1)\,{\mathbbm 1}~,  
 \eea 
 \end{subequations}
where the quantum number $s$ is called spin. Its possible values 
in different representations are 
$s= 0, 1/2, 1, 3/2 \dots$. 
The massless representations are characterised by the condition $P^aP_a =0$. 
For the physically interesting massless representations, it holds that 
\bea
{\mathbb W}_a = \l P_a~,
\label{1.5}
\eea
where the parameter $\l$ determines the representation and is called the helicity. Its possible values are $0, \pm\hf, \pm 1, $ and so on. The parameter $|\l |$ is called the spin of a massless particle.

In this paper we present a generalisation of Wigner's equation \eqref{1.5}
to five and higher dimensions.


\section{Unitary representations of $\sISO_0(4,1)$}

The five-dimensional analogue of \eqref{PLv} is the  Pauli-Lubanski tensor
\begin{equation}
\mathbb{W}^{ab} = \frac{1}{2}\ve^{abcde}J_{cd}P_e~.
\label{2.1}
\end{equation}
It is translationally invariant, 
\bea
\big[  {\mathbb W}_{ab} , P_c \big] =0~,
\eea
and possesses the following properties:
\begin{subequations}
\bea
\mathbb{W}_{ab} P^b &=& 0~, \\
\big[\mathbb{W}_{ab}, J_{cd}\big] &=& \ri \eta_{ac}\mathbb{W}_{bd} 
- \ri \eta_{ad}\mathbb{W}_{bc}  - \ri \eta_{bc}\mathbb{W}_{ad} + \ri \eta_{bd}\mathbb{W}_{ac}~, \\
 \big[\mathbb{W}_{ab}, \mathbb{W}_{cd}\big] 
 &=& \ri\ve_{acdfg} {\mathbb{W}_b}^f  P^{g} - \ri \ve_{bcdfg} {\mathbb{W}_a}^f P^{g}~.
\eea
\end{subequations}
Making use of ${\mathbb W}_{ab}$ allows one to construct two Casimir operators, which are 
\begin{equation}
\mathbb{W}_{ab} \mathbb{W}^{ab}~,  \qquad 
\mathbb{H} := \mathbb{W}^{ab}J_{ab}~.
\label{2.4}
\end{equation}


\subsection{Irreducible massive representations} 

The irreducible massive representations of the Poincar\'e group $\sISO_0(4,1)$
are characterised by two conditions
\begin{subequations}\label{2.5}
\bea
 \frac{1}{8} \Big( \mathbb{W}^{ab}\mathbb{W}_{ab} + m \mathbb{H} \Big) &=& 
 m^2 s_1(s_1 + 1)\mathbbm{1} ~,  \\
\frac{1}{8} \Big( \mathbb{W}^{ab}\mathbb{W}_{ab} - m\mathbb{H} \Big) &=& 
m^2 s_2(s_2 + 1)\mathbbm{1} ~,
\eea
\end{subequations}
in addition to \eqref{1.4a}.
Here $s_1$ and $s_2$ are two spin values corresponding to the two $\sSU(2)$ subgroups of the universal covering group 
$\mathsf{Spin}(4) \cong \sSU(2) \times \sSU(2)$
of the little group.\footnote{The equations \eqref{2.5} were independently derived during the academic year 1992-93 by Arkady Segal and David Zinger, who were undergraduates at Tomsk State University at the time.}


\subsection{Irreducible massless representations} \label{section2.2}

It turns out that all irreducible massless representations of 
$\sISO_0(4,1)$ with a finite spin are characterised by the condition 
\bea
\varepsilon_{abcde}P^{c}\mathbb{W}^{de} =0 \quad \Longleftrightarrow \quad
P^{[a} {\mathbb W}^{bc]} =0~.
\label{main}
\eea
Both Casimir operators \eqref{2.4} are equal to zero in these representations, 
$\mathbb{W}_{ab} \mathbb{W}^{ab}=0$  and   
$ \mathbb{W}^{ab}J_{ab} =0$.

Let $ \ket{p, \sigma} $ be an orthonormal basis in the Hilbert space of one-particle states, where $p^a $ 
denotes the momentum of a particle, $ P^a \ket{p, \sigma} = p^a  \ket{p, \sigma} $,
and $\s$ stands for the spin degrees of freedom.
For a massless particle, we choose as our standard 5-momentum $k^{a} = (E,0,0,0,E)$. On this eigenstate:
\begin{equation}
\mathbb{W}^{ab} \ket{k, \sigma} = \frac{1}{2}\varepsilon^{abcde}J_{cd}P_{e} \ket{k, 
\sigma} = \frac{E}{2} \left(\varepsilon^{abcd4}J_{cd} - \varepsilon^{abcd0}J_{cd} \right) \ket{k, \sigma}~.
\end{equation}
Running through the elements of $\mathbb{W}^{ab}$, one finds:
\begin{equation} \label{eq:eq5}
\begin{split}
\mathbb{W}^{01} = \mathbb{W}^{41} = -EJ_{23} ~,\qquad & \mathbb{W}^{12} = E(J_{30} +J_{34}) ~,\\
\mathbb{W}^{02} = \mathbb{W}^{42} = -EJ_{31} ~,\qquad & \mathbb{W}^{23} = E(J_{10} +J_{14}) ~,\\
\mathbb{W}^{03} = \mathbb{W}^{43} = -EJ_{12}~, \qquad & \mathbb{W}^{31} = E(J_{20} +J_{24}) ~,\\
\mathbb{W}^{04} = & \ 0 ~.
\end{split} 
\end{equation}
If we rescale these generators and define:
\begin{equation}\label{eq:eq7}
\begin{split}
\mathbb{R}_1  \equiv \frac{1}{E}\mathbb{W}^{23}~, \qquad \mathbb{R}_{2} \equiv &  \frac{1}{E}\mathbb{W}^{31}~,  \qquad \mathbb{R}_{3} \equiv \frac{1}{E}\mathbb{W}^{12} ~,\\
\qquad \cJ_{i} \equiv &-  \frac{1}{E} \mathbb{W}^{0i} ~,
\end{split}
\end{equation}
then these new operators satisfy:
\begin{equation}
\big[\cJ_{i}, \cJ_{j} \big] = \ri\varepsilon_{ijk}\cJ_k~, \qquad 
\big[\cJ_i , \mathbb{R}_j \big] = \ri\varepsilon_{ijk} \mathbb{R}_k~ , 
\qquad \big[\mathbb{R}_i , \mathbb{R}_j \big] = 0~.
\end{equation}
These are the commutation relations for the three-dimensional Euclidean algebra, $\mathfrak{iso}(3)$. The irreducible unitary representations of $\mathfrak{iso}(3)$ are labelled by a continuous parameter $\mu^2$, corresponding to the value the Casimir operator $\mathbb{R}^{i} \mathbb{R}_{i}$ takes. Since $\mathbb{R}_{i}$ commute among themselves the operators can be simultaneously diagonalised, and the eigenvectors $\ket{r_{i}}$ taken as a basis. However the only restriction on these is that $r_i r^i = \mu^2$, which for non-zero $\mu^2$ permits a continuous basis and is thus an infinite dimensional representation. Because we want only finite-dimensional representations, we must take:
\begin{equation} \label{eq:eq6}
\mu^2 = 0 \quad \Longrightarrow \quad \mathbb{R}_{i} = 0 \quad \Longleftrightarrow \quad J_{0i} = -J_{4i}~.
\end{equation}
We are therefore restricted to those representations in which the translation component is trivial, and so only the generators $\cJ_i$ remain, which generate the algebra $\mathfrak{so}(3)$. The algebra of the little group on massless representations is thus $\mathfrak{so}(3) $ which is isomorphic to $\mathfrak{su}(2)$. 
As stated previously, the irreducible representations of $\mathfrak{su}(2)$ are labelled  by a non-negative (half) integer $s$ and have a single Casimir operator $\cJ^i \cJ_i$ which takes the value $s(s+1)\mathbbm{1}$. This analysis leads to \eqref{main}.

The spin value of a massless representation can still be found using a `spin' operator.
The following relation holds on massless representations:
\begin{equation}
 \mathbb{S}_a :=- \frac{1}{4} \varepsilon_{abcde} J^{bc} \mathbb{W}^{de}  
 = \cJ^2 P_{a} = s(s+1)P_a~,
 \label{2.12}
\end{equation}
where $\cJ^2 = \cJ^{i}\cJ_{i}$ is the Casimir operator for the $\mathfrak{so}(3)$ generators in \eqref{eq:eq7}. The parameter $s$ is the spin of a massless particle.
Its possible values in different representations are 
$s= 0, 1/2, 1, $ and so on. Equation \eqref{2.12} naturally holds for massless  spinor and vector fields \cite{Pindur}.

In general, the operator ${\mathbb S}_a$ is not translationally invariant,
\begin{equation}
\big[ \mathbb{S}_b, P_{a} \big] = \frac{\ri}{2}\varepsilon_{abcde}P^{c}\mathbb{W}^{de} ~.
\end{equation}
It is only for the massless representations with finite spin that  the quantity on the right vanishes so that the spin operator commutes with the momentum operators. 
Equation \eqref{2.12} is the five-dimensional analogue of the operator equation \eqref{1.5}. Its consistency condition is \eqref{main}.\footnote{The consistency condition for \eqref{1.5} is  $P^{[a} {\mathbb W}^{b]} =0$, which is the four-dimensional counterpart of \eqref{main}.}


\section{Generalisations} 

The results of section \ref{section2.2} can be generalised to $d>5$ dimensions. 
The Pauli-Lubanski tensor \eqref{2.1} turns into
\bea
\mathbb{W}^{a_1 \dots a_{d-3}} = 
\frac{1}{2}
\ve^{a_1 \dots a_{d-3} bc e}J_{bc}P_e~.
\label{3.1}
\eea
The condition \eqref{main} is replaced with
\bea
P^{[a} {\mathbb W}^{b_1 \dots b_{d-3}]} =0~.
\label{maind}
\eea
This equation is very similar to another that has appeared in the literature
using the considerations of conformal invariance 
\cite{Bracken,BrackenJ,Siegel,SiegelZ}. One readily checks that \eqref{maind} is equivalent to 
\bea
J_{ab} P^2 + 2 J_{c[a} P_{b]} P^c =0 \quad \implies \quad 
J_{c[a} P_{b]} P^c =0 ~.
\label{maind2}
\eea
The latter is solved on the momentum eigenstates 
by $J_{ab} p^b \propto p_a$, which is of the form considered
in \cite{Bracken,BrackenJ,Siegel,SiegelZ}.\footnote{We are grateful to Warren Siegel for useful comments.}

Equation \eqref{maind} characterises all irreducible massless representations of 
$\sISO_0(d-1,1)$ with a finite (discrete) spin. Finally, the spin equation \eqref{2.12}
turns into 
\bea
 \mathbb{S}_a := \frac{(-1)^d}{ 2(d-3)!} \varepsilon_{abc e_1 \dots e_{d-3}} J^{bc} \mathbb{W}^{e_1 \dots e_{d-3}}  
 = \cJ^2 P_{a} ~,
\eea
where $\cJ^2 = \hf \cJ^{ij} \cJ_{ij} $ is the quadratic Casimir operator of  the algebra $\mathfrak{so}(d-2)$, with $i,j=1, \dots , d-2$.
For every irreducible massless representation of 
$\sISO_0(d-1,1)$ with a finite spin, it holds that $\cJ^2 \propto {\mathbbm 1}$.

We can extend this further to higher-order Casimir operators of $\mathfrak{so}(d-2)$. 
As a generalisation of \eqref{3.1}, we introduce the $n^{\text{th}}$ Pauli-Lubanski tensor 
\begin{equation}
{\mathbb{W}^{(n)}}_{a_1 \ldots a_{d-2n-1}}  =
 \frac{1}{2^n}
  \varepsilon_{a_1 \ldots a_d}J^{a_{d-2n} a_{d-2n+1}} \ldots J^{a_{d-2} a_{d-1}}P^{a_d}~,
  \qquad 1\leq n \leq \lfloor \frac{d-2}{2} \rfloor
\end{equation}
which is order $n$ in the Lorentz generators (the operator \eqref{3.1} coincides with 
${\mathbb{W}^{(1)}}$).\footnote{In the massless case, all Casimir operators of the Poincar\'e group $({\mathbb{W}^{(n)}}_{a_1 \ldots a_{d-2n-1}})^2 $
 vanish, and so does the scalar operator ${\mathbb{W}^{(\frac{d-1}{2})}}$, which is 
 defined when $d$ is odd.}
Then higher-order  spin operators can be defined as
\begin{equation}
{\mathbb{S}^{(n)}}_{a_1} = \frac{(-1)^d}{2(d-2n-1)!}\varepsilon_{a_1 \ldots a_d}J^{a_2 a_3} \ldots J^{a_{2n} a_{2n+1}} {\mathbb{W}^{(n) a_{2n+2} \ldots a_d}}~,
\end{equation}
which are order $2n$ in the Lorentz generators. 
Using the fact that $J^{a \, 0} = J^{a \, d-1}$
in the frame with a standard $d$-momentum   $k^{a} = (E,0,\dots,0,E)$,
one can show that 
\begin{equation}
{\mathbb{S}^{(n)}}_{a} = \mathcal{C}^{(n)} P_{a}
\label{3.7}
\end{equation}
where $\mathcal{C}^{(n)}$ is a an order $2n$ Casimir operator for $\mathfrak{so}(d-2)$ defined by
\begin{align}
\begin{split}
\mathcal{C}^{(n)} & = \frac{-1}{2^{n+1}(d-2n-2)!} \varepsilon_{0 \, i_1 \ldots i_{d-2} \, d-1} \varepsilon^{0 \, j_{1} \ldots j_{2n} i_{2n+1} \ldots i_{d-2} \, d-1} \times  \\ 
& \qquad \qquad \qquad \qquad \qquad \qquad \qquad  J^{i_1 i_2} \ldots J^{i_{2n-1} i_{2n}}J_{j_1 j_2} \ldots J_{j_{2n-1} j_{2n}} 
\label{3.8}\\
& = \frac{(2n)!}{2^{n+1}} J^{i_1 i_2} \ldots J^{i_{2n-1} i_{2n}}J_{[i_1 i_2} \ldots J_{i_{2n-1} i_{2n]}}~.
\end{split}
\end{align}
If $d$ is odd, the operators \eqref{3.8} can be constructed up to $n = \frac{d-3}{2}$ (the order $n=\frac{d-1}{2}$ Pauli-Lubanski tensor is a scalar). If $d$ is even, 
it suffices to restrict $n$ to run from 1 
to $n = \frac{d-4}{2}$, since the Pauli-Lubanski tensor of order $\frac{d-2}{2}$,
\begin{equation}
\mathbb{W}^{(\frac{d}{2} -1)}{}_{a_1} = 
\frac{1}{2^{\hf d -1}} 
\varepsilon_{a_1 \ldots a_d}J^{a_2 a_3} \ldots J^{a_{d-2} a_{d-1}}P^{a_d}~,
\end{equation}
is itself a `spin operator' with the property
\begin{equation}
\mathbb{W}^{(\frac{d}{2} -1)} {}_{a} = \L^{(\frac{d}{2} -1)}P_{a}~, 
\label{3.10}
\end{equation}
where
\begin{equation}
\L^{(\frac{d}{2} -1)} = -
\frac{1}{2^{\hf d -1}} 
\varepsilon_{0 i_1 \ldots i_{d-2} d-1} J^{i_1 i_2} \ldots J^{i_{d-3} i_{d-2}}
\end{equation}
Note that the $d=4$ case corresponds to \eqref{1.5}. In the $d=6$ case, the equation 
\eqref{3.10} was pointed out in \cite{MRT}.

For every irreducible massless representation of 
$\sISO_0(d-1,1)$ with a finite spin, the operator $\cC^{(n)}$ in \eqref{3.7} 
is a multiple of the identity operator, $\cC^{(n)}  \propto {\mathbbm 1}$. 
Then the translational invariance of the equations \eqref{3.7} implies 
  \eqref{maind} and the  relation
\bea
{\mathbb W}^{(n-1)}{}_{a_1 a_2 b_1 \dots b_{d-2n-1} } 
{\mathbb W}^{(n)}{}^{b_1 \dots b_{d-2n-1} } =0~.
\eea

It is possible to derive a five-dimensional analogue of the operator equation 
defining the $\cN=1$ superhelicity $\k$  in four dimensions \cite{BK}. 
The latter has the form\footnote{In the supersymmetric case, the conventions of \cite{BK} are used, 
in particular the Levi-Civita tensor $\ve_{abcd}$ is normalised by $\ve_{0123}=-1$.}  
\bea
{\mathbb L}_a = \Big( \k +\frac 14 \Big) P_a~, 
\label{3.4}
\eea
where the operator ${\mathbb L}_a$ is defined by 
\bea
{\mathbb L}_a = {\mathbb W}_a - \frac{1}{16} (\tilde \s_a)^{\ad \a} \big[ Q_\a , \bar Q_\ad\big] ~.
\label{3.5}
\eea
The fundamental properties of the operator ${\mathbb L}_a $
(the latter differs from the supersymmetric Pauli-Lubanksi vector \cite{SS})
are that it is translationally invariant and commutes with the supercharges $Q_\a$ and $\bar Q_\ad$ in the massless representations of the $\cN=1$ super-Poincar\'e group.\footnote{The irreducible massless
 representation of superhelicity $\k$ is the direct sum of two irreducible massless Poincar\'e representations corresponding to the helicity values $\k$ and $\k+\hf$.} 
The superhelicity operator \eqref{3.5} was generalised to higher dimensions
in \cite{PZ,AMT}. Generalisations of \eqref{3.4} to five and higher dimensions will be discussed elsewhere.
\\

\noindent
{\bf Acknowledgements:}\\
We thank Warren Siegel for pointing out important references, and Michael Ponds for comments on the manuscript. 
SMK is grateful to Ioseph Buchbinder for email correspondence, and to Arkady Segal for discussions. The work of SMK work is supported in part by the Australian 
Research Council, project No. DP200101944.


\begin{footnotesize}

\end{footnotesize}


\end{document}